# Scoping Electronic Communication Privacy Rules: Data, Services and Values

by **Joris van Hoboken and Frederik Zuiderveen Borgesius***

**Abstract:** We use electronic communication networks for more than simply traditional telecommunications: we access the news, buy goods online, file our taxes, contribute to public debate, and more. As a result, a wider array of privacy interests is implicated for users of electronic communications networks and services. This development calls into question the scope of electronic communications privacy rules. This paper analyses the scope of these rules, taking into account the rationale and the historic background of the European electronic communications privacy framework. We develop a framework for analysing the scope of electronic communications privacy rules using three approaches: (i) a service-centric approach, (ii) a data-centric approach, and (iii) a value-centric approach. We discuss the strengths and weaknesses of each approach. The current e-Privacy Directive contains a complex blend of the three approaches, which does not seem to be based on a thorough analysis of their strengths and weaknesses. The upcoming review of the directive announced by the European Commission provides an opportunity to improve the scoping of the rules.

Recommended citation: Joris van Hoboken and Frederik Zuiderveen Borgesius, Scoping Electronic Communication Privacy Rules: Data, Services and Values, 6 (2015) JIPITEC 198, para 1.

## A. Introduction

1 Sector-specific frameworks for electronic communications privacy, such as the European Union e-Privacy Directive,[1] have their historical roots in the sector-specific rules for public telecommunications networks, used for one-to-one voice communications. Nowadays, we use electronic communications networks for a wide variety of purposes beyond traditional telecommunications, including commerce, work, social interaction, media access, and interaction with government. The privacy interests of users engaged in these different activities go far beyond the interests protected in the current e-Privacy Directive. Therefore, the scope of the electronic communications privacy rules should be reassessed.

2 Currently, the e-Privacy Directive leaves considerable gaps in user protection; for instance because the rules for location and traffic data do not apply to new players in the electronic communications sector. The EU lawmaker has not systematically addressed user privacy interests related to access to online content, interactive media, and the wide variety of opportunities offered by networked communications. In 2015, the European Commission announced a review of the e-Privacy Directive.[2] In such a review, the question regarding the scope of the rules will be important.

---

1 Council Directive 2002/58/EC of 12 July 2002 on the processing of personal data and the protection of privacy in the electronic communications sector, as amended by Council Directive 2006/24/EC and Council Directive 2009/136/EC (e-Privacy Directive).

2 See European Commission, Communication of 6 May 2015 on A Digital Single Market for Europe, COM (2015) 192 final, p. 13.

**3** In this paper, we discuss three different approaches to scoping electronic communications privacy rules. Distinguishing these three approaches can aid in reaching informed decisions regarding scoping the rules. The three approaches are: (i) a service-centric approach, (ii) a data-centric approach, and (iii) a value-centric approach. (i) In a service-centric approach, the scope of the rules is delineated on the basis of different services. (ii) A data-centric approach protects privacy interests of users through the proxy of setting rules for processing types of personal data. (iii) A value-centric approach determines the scope of the rules based on the user's privacy interests at stake when using electronic communications networks.

**4** We do not argue that one of the approaches is better than another – each approach has strengths and weaknesses. We provide the distinction between the three approaches as an analytical tool to assist in structuring discussions about scoping electronic privacy rules.

**5** The article is structured as follows. In section two, we discuss the background and the scope of the main provisions of the e-Privacy Directive. The service-centric, data-centric, and value-centric approaches are outlined in sections three, four, and five respectively. The final section concludes that the European lawmaker should be aware of the strengths and weaknesses of the different approaches involved in scoping electronic communications privacy rules.

## B. The e-Privacy Directive: background and current scope

**6** In 1990, the European Commission presented a proposal for a Data Protection Directive with the aim to harmonise data privacy regimes to foster the European single market. After long and heated debates, the Data Protection Directive was finally adopted in 1995.[3] Additionally in 1990, the European Commission presented a proposal for a telecommunications privacy directive. The European Commission was planning to adopt the telecommunications privacy directive at the same time as the Data Protection Directive,[4] but it took until 1997.[5]

**7** The general Data Protection Directive and the e-Privacy Directive are internal market harmonisation instruments.[6] The Data Protection Directive's dual aim is to provide a high level of data protection across the member states, and to ensure that personal data can flow across borders within Europe, uninhibited by differences in data privacy laws.[7] The e-Privacy Directive has a similar dual aim, for the electronic communications sector.[8]

**8** In 2002, the 1997 telecommunications privacy directive was replaced by the e-Privacy Directive, officially the "Directive concerning the processing of personal data and the protection of privacy in the electronic communications sector." This e-Privacy Directive was intended to be more in line with new technologies.[9] In 2009, the e-Privacy Directive was updated by the Citizens' Rights Directive.[10] Some of the key changes were the introduction of a consent requirement for tracking cookies and similar files, and an obligation to report data breaches.[11]

**9** A few years earlier, in 2006, the European lawmaker had adopted the Data Retention Directive as an amendment to the e-Privacy Directive.[12] The Data Retention Directive obliged member states to require retention of electronic communications data by

---

related service providers for a period of 6-24 months, to enable government agencies to access these data. In 2014, the Court of Justice of the European Union declared this directive invalid.[13]

10   The e-Privacy Directive is a sector-specific regulatory instrument, adopted as part of the EU regulatory package for the telecommunications sector.[14] The directive's full title illustrates the goal of sector-specificity; the directive concerns "the processing of personal data and the protection of privacy in the electronic communications sector" [emphasis added].

11   Most of the provisions in the e-Privacy Directive contain rules applicable to "providers of publicly available electronic communications services", and "providers of public communications networks".[15] The scope of these e-Privacy Directive provisions is thus narrower than the scope of the general Data Protection Directive. The latter applies, in short, as soon as "personal data" are processed, regardless of the sector (with exceptions).[16]

12   For its material scope, the e-Privacy Directive partly relies on the definitions in the Framework Directive for electronic communications networks and services.[17] The resulting scope is not always clear, and is suboptimal from the perspective of protecting users' electronic communications privacy. As discussed below, there are many over-the-top services that are, from a user perspective, functionally equivalent to "publicly available electronic communications services" – but those over-the-top services do not fall within that definition.

13   An electronic communications network is defined in the Framework Directive as: "transmission systems and, where applicable, switching or routing equipment and other resources, including network elements which are not active, which permit the conveyance of signals by wire, radio, optical or other electromagnetic means, including satellite networks, fixed (circuit- and packet-switched, including Internet) and mobile terrestrial networks, electricity cable systems, to the extent that they are used for the purpose of transmitting signals, networks used for radio and television broadcasting, and cable television networks, irrespective of the type of information conveyed."[18]

14   The Framework Directive defines an "electronic communications service" as "a service normally provided for remuneration which consists wholly or mainly in the conveyance of signals on electronic communications networks, including telecommunications services and transmission services in networks used for broadcasting, but exclude services providing, or exercising editorial control over, content transmitted using electronic communications networks and services; it does not include information society services (…) which do not consist wholly or mainly in the conveyance of signals on electronic communications networks."[19]

15   An electronic communications service is, in short, a service that consists wholly or primarily in the conveyance of signals on electronic communications networks. This implies for instance, that the e-Privacy Directive is not applicable to voice over IP (VoIP) software services such as Skype, even though for users such services may be functionally equivalent to regulated services such as telephony. For personal data processing by services outside of the scope of the e-Privacy Directive, the general rules in the Data Protection Directive still apply.[20] From a user's privacy perspective, this difference in legal treatment does not make sense. In practice, individuals may not even be aware whether they are making a call through an electronic communications service or through a VoIP service.

16   Furthermore, as established in article 3, generally the e-Privacy Directive only applies to publicly available services and networks. This restriction has led to much debate. The Article 29 Working Party, in which European Data Protection Authorities cooperate, noted in 2008 that the distinction between private and public networks and services is difficult to draw: "Services are increasingly becoming a mixture of private and public elements and it is often difficult for regulators and for stakeholders alike to determine whether the e-Privacy Directive applies in a given situation. For example, is the provision of internet access to 30.000 students a public electronic communication system or a private one? What if the same access is provided by a multinational company, to tens of thousands of employees? What if it is provided by a cybercafé?"[21]

---

13   Court of Justice of the EU 8 April 2014, C-293/12 (Digital Rights Ireland).

14   See recital 4 of the e-Privacy Directive.

15   See infra Section 3 for more discussion.

16   See article 1(1) of the Data Protection Directive. Some parts of the public sector are outside the scope of the Directive (see article 3(2) and article 13). Some data processing practices in the private sector are also exempted, for purely personal purposes (article 3(2). There are also exemptions for the processing for journalistic purposes (article 9).

17   Council Directive 2002/21 of 7 March 2002 on a common regulatory framework for electronic communications networks and services as amended by Directive 2009/140/EC and Regulation 544/2009 (Framework Directive).

18   Article 2(a) of the Framework Directive.

19   Article 2(c) of the Framework Directive.

20   Recital 10 of the e-Privacy Directive.

21   Article 29 Working Party, 'Opinion 2/2008 on the review of the Directive 2002/58/EC on privacy and electronic com-

**17** Article 4 deals with the security of processing, and contains notification obligations regarding data breaches.[22] The security requirements and the data breach notification obligation in article 4 only apply to providers of publicly available electronic communications services.[23]

**18** The e-Privacy Directive's specific regime for traffic and location data in article 5, 6 and 9 is roughly as follows. Unless a specified exception applies, consent of the user or subscriber is required for the processing of traffic and location data by regulated services. Traffic data, sometimes called metadata, are "any data processed for the purpose of the conveyance of a communication on an electronic communications network or for the billing thereof".[24] Examples of traffic data are the time of a communication, and the addressing information of those involved in a communication, such as the email address or IP address used to access the internet.[25] With modern communication technology, the line between traffic data and communications content has become increasingly blurred. For instance, the subject line of an email message could be seen as traffic data or as communications content. Monitoring communications traffic data over time can provide a detailed picture of individuals' lives.[26]

**19** Location data are data "indicating the geographic position of the terminal equipment of a user of a publicly available electronic communications service".[27] Location data can be sensitive.[28] For example, a phone's location data can disclose visits to a hospital, church, or mosque, or the location of one's bed.

**20** Reflecting the telecommunications service background of the e-Privacy Directive, article 7 assumes that "subscribers" receive itemised bills, and grants them the right to receive non-itemised bills.[29] A subscriber is defined as "any natural person or legal entity who or which is party to a contract with the provider of publicly available electronic communications services for the supply of such services".[30]

**21** Article 8 concerns privacy interests related to calling line identification on a per-call basis. Under article 11, subscribers must be able, by request to the provider of the publicly available electronic communications service, to stop forwarded calls being passed on to them. The scope of article 8 and 11 is limited to providers of publicly available electronic communications services and networks. Article 8 and 11 apply to "calls". A call refers, in brief, to voice telephony.[31]

**22** Article 5(1) emphasises member states' positive obligations regarding communications confidentiality.[32] Article 5(1) can be summarised as follows: member states must ensure the confidentiality of communications and the related traffic data by means of publicly available electronic communications services. In particular, member states must prohibit tapping, storage or other kinds of surveillance of communications, without the consent of the users or other legal authorisation.

**23** The scope of these positive obligations for member states is subject to debate. If an internet access provider employs deep packet inspection

---

munications (e-Privacy Directive)' (WP150), Brussels, 15 May 2008, p. 4. See also: European Commission, 'ePrivacy Directive: assessment of transposition, effectiveness and compatibility with proposed Data Protection Regulation', Final Report (a study prepared for the European Commission DG Communications Networks, Content & Technology by Time.Lex and Spark legal network and consultancy ltd, 10 June 2015) http://ec.europa.eu/newsroom/dae/document.cfm?doc_id=9962 (accessed 15 November 2015), p. 24-32.

22    Article 4(3)-4(5) of the e-Privacy Directive. See also Recital 61 of the Citizens' Rights Directive.

23    Article 4(1) of the e-Privacy Directive.

24    Article 2(b) of the e-Privacy Directive.

25    Recital 15 of the e-Privacy Directive.

26    See e.g. B.J Koops & J.M. Smits, 'Verkeersgegevens en artikel 13 Grondwet. Een technische en juridische analyse van het onderscheid tussen verkeersgegevens en inhoud van communicatie' [Traffic data and article 13 of the Constitution. Technical and legal analysis of the distinction between traffic data and communications content], Wolf Legal Publishers 2014; P. Breyer, 'Telecommunications data retention and human rights: the compatibility of blanket traffic data retention with the ECHR', European Law Journal 2014, Vol. 11, No. 3, pp. 365-375; E. Felten, 'Written Testimony, Committee on the Judiciary Hearing on Continued Oversight of the Foreign Intelligence Surveillance Act', www.cs.princeton.edu/~felten/testimony-2013-10-02.pdf (accessed 15 November 2015); J. Mayer, & P. Mutchler, 'MetaPhone: The Sensitivity of Telephone Metadata' 2014, webpolicy.org/2014/03/12/metaphone-the-sensitivity-of-telephone-metadata/ (accessed 15 November 2015); H. de Zwart, 'How your innocent smartphone passes on almost your entire life to the secret service' 2014, www.bof.nl/2014/07/30/how-your-innocent-smartphone-passes-on-almost-your-entire-life-to-the-secret-service/ (accessed 15 November 2015); J.C. Fischer, Communications Network Traffic Data - Technical and Legal Aspects

(PhD thesis University of Eindhoven), Academic version 2010 http://alexandria.tue.nl/extra2/689860.pdf accessed 15 November 2015.

27    Article 9 of the e-Privacy Directive.

28    Article 29 Working Party, 'Opinion 13/2011 on Geolocation services on smart mobile devices' (WP 185) 16 May 2011, p. 7.

29    Article 7(1) of the e-Privacy Directive.

30    Article 2(k) of the Framework Directive.

31    See article 2(s) of the Framework Directive.

32    See W. Steenbruggen, 'Publieke dimensies van privé-communicatie: een onderzoek naar de verantwoordelijkheid van de overheid bij de bescherming van vertrouwelijke communicatie in het digitale tijdperk' [Public dimensions of private communication: an investigation into the responsibility of the government in the protection of confidential communications in the digital age], PhD thesis University of Amsterdam, Cramwinkel 2009, p. 176, p. 356.

to analyse people's internet use, including email communication, article 5(1) applies, since internet access providers are publicly available electronic communications services.

24 However, the broad formulation of article 5(1) could imply that member states' positive obligations extend to services involved in electronic communications that are not publicly available electronic communications services in the strict sense of the e-Privacy Directive. Thus, member states would have to ensure that nobody interferes with the confidentiality of communications and related traffic data flowing over public communications networks[33] A similar general positive obligation could be based on the fundamental right to private life and private correspondence in Article 8 of the European Convention on Human Rights and Article 7 of the EU Charter of Fundamental Rights.

25 Web browsing and using online video services also fall within the legal definition of communication in the e-Privacy Directive.[34] Monitoring people's web browsing is thus only allowed after their consent, as member states must prohibit "interception or surveillance of communications and the related traffic data by persons other than users, without the consent of the users concerned".[35] The European Data Protection Supervisor says that article 5(1) does not only apply to electronic communication service providers and networks, but has a broader scope.[36]

26 The rules for spam and cookies have a different scope than the majority of the other rules in the e-Privacy Directive. In short, article 11 only allows sending marketing emails after the receiver's prior consent is obtained (subject to exceptions for mail to existing customers).

27 Article 5(3) applies to anyone that stores or accesses information, such as a cookie, on a user's device, including if no personal data are involved. Article 5(3) is hotly debated, because it applies to tracking of internet users through cookies for online marketing. [37] The preamble shows that article 5(3) aims to protect the user's device and its contents against unauthorised access: "Terminal equipment of users of electronic communications networks and any information stored on such equipment are part of the private sphere of the users requiring protection under the European Convention for the Protection of Human Rights and Fundamental Freedoms."[38] The provision applies, for instance, to apps that access information on a user's smartphone, such as location data or a user's contact list.[39] Article 5(3) also protects the user against parties that want to store spyware on a user's device, without the user's knowledge. While article 5(3) does address privacy interests related to the use of electronic communication networks, its scope is atypical. The parties placing cookies or other information on user devices are not the parties that are generally regulated by the e-Privacy Directive.

28 In sum, the majority of the e-Privacy Directive's provisions only apply to publicly available communications networks or services. In the next section, we discuss the strengths and weaknesses of this service-centric approach to scoping electronic communications privacy rules.

## C. A service-centric approach

29 In a service-centric approach to develop the scope of electronic communications privacy rules, the scope of the rules is delineated on the basis of different services. In brief, such rules only apply to certain types of companies operating in relevant electronic communications markets.

**30** As suggested earlier, the e-Privacy Directive largely uses a service-centric approach. The main reason for this specific scoping is the directive's background as part of the regulatory framework for electronic communications markets.[40] A central feature of this regulatory framework is the recognition of the specific market characteristics of electronic communications networks and services, and their value for users and society. The framework recognises the particular market entry dynamics and network effects in the telecommunications industry. The framework aims to foster competition between relevant services, while providing for interconnection and interoperability of networks and services.[41]

**31** Electronic communications networks and services constitute the electronic communications infrastructure, whereas over-the-top services merely use such infrastructure. Regulating the privacy conditions of infrastructure services also affects the privacy conditions of services that become available for use over such infrastructure, including over-the-top services, such as communications software. Hence, the focus on electronic communications services and networks involved in transmission activities can be defended on the basis of the infrastructural nature of these services for electronic communications. These services can have a more significant impact on communications privacy than other services that do not qualify as infrastructure.

**32** A particular strength of a service-centric approach is that – if done right – it can be reasonably clear for a company whether it has to comply with a rule. The company must simply assess whether it is a "provider of a publicly available electronic communications service", or a "provider of a public communications network". Hence, in principle a service-centric approach can lead to rules with a relatively clear scope.

**33** The key weakness of a service-centric approach is that such an approach can lead to – from a user perspective – arbitrary differences between protections for different but functionally equivalent services. For example, the e-Privacy Directive's rules for traffic and location data only apply to "providers of publicly available electronic communications services", and to "providers of public communications networks." However many companies, such as advertising networks (a type of online marketing company) and providers of smart phone apps, process data of a more sensitive nature than telecommunications providers. However, ad networks and apps providers are not subject to the e-Privacy Directive's rules for traffic and location data. Such companies are subject to the Data Protection Directive as far as they process personal data.

**34** General data protection law is less stringent and less specific than the e-Privacy Directive's regime for traffic and location data. For instance, under the general Data Protection Directive, a data controller can rely on several legal bases for processing personal data – not only on the data subject's consent. An advertising network could, for instance, try to argue that for processing location data it has a legitimate interest that overrides the data subject's fundamental rights, and that therefore, it may process the data without the data subject's consent.[42] From a user's perspective, it is not logical that the rules are less strict when location data are in the hands of an advertising network, than when they are in the hands of an internet access provider.[43]

**35** For other provisions in the e-Privacy Directive, such as the data breach notification requirement, the restriction to providers of publicly available electronic communications services appears also without merit. The e-Privacy Directive requires an internet access provider (a provider of a publicly available electronic communications service) to notify the authorities when an employee loses a laptop with customer data. But the e-Privacy Directive does not require a webmail provider, an online bank, or an online pharmacy to notify users and authorities of data breaches.[44]

**36** Before the 2009 amendments to the e-Privacy Directive were adopted, there was ample discussion regarding the scope of the data breach notification requirements. The Article 29 Working Party, the European Data Protection Supervisor, and the European Parliament were in favour of extending the scope of the notification requirements to, at least, all providers of information society services.[45]

---

40   See C. Schnabel, 'Privacy and Data Protection in Electronic Communications Law', in C. Koenig, et al. (eds), EC Competition and Telecommunications Law, Kluwer Law International 2009, pp. 509-568, p. 520-522.

41   See e.g. P. Alexiadis & M. Cave, 'Regulation and Competition Law in Telecommunications and Other Network Industries', in: R. Baldwin, M. Cave & M. Lodge (eds.), The Oxford handbook of Regulation, Oxford University Press 2010.

42   See article 7(f) of the Data Protection Directive. See F.J. Zuiderveen Borgesius, 'Personal data processing for behavioural targeting: which legal basis?', International Data Privacy Law, doi: 10.1093/idpl/ipv011, 2015.

43   See also F.J. Zuiderveen Borgesius, 'Improving Privacy Protection in the area of Behavioural Targeting', Kluwer Law International 2015, pp. 282-283.

44   See F.J. Zuiderveen Borgesius, 'De Meldplicht Voor Datalekken in De Telecommunicatiewet' [The data breach notification requirement in the Dutch Telecommunications Act], Computerrecht 2011, No. 4, pp. 209-218; A. Arnbak, 'Securing Private Communications' (PhD thesis University of Amsterdam, academic version), http://hdl.handle.net/11245/1.492674 (accessed 15 November 2015), p. 48-49.

45   Article 29 Working Party, 'Opinion 1/2009 on the proposals amending Directive 2002/58/EC on privacy and electronic

The European Commission did not follow that suggestion. However, in the 2012 proposal for a Data Protection Regulation, the Commission did introduce a data breach notification requirement.[46] If that proposal were adopted, it would be difficult to see why sector-specific data breach rules in the e-Privacy Directive would still be needed.

37  Indeed, recently the European Commission suggested that the narrow scope of the e-Privacy Directive should be reassessed:

> *Special rules apply to electronic communications services (e-Privacy Directive) which may need to be reassessed once the general EU rules on data protection are agreed, particularly since most of the articles of the current e-Privacy Directive apply only to providers of electronic communications services, i.e. traditional telecoms companies. Information society service providers using the Internet to provide communication services are thus generally excluded from its scope.[47]*

38  In sum, the main strength of the service-centric approach to scoping electronic communications privacy rules is the possibility of clear scoping. Another argument in favour of a service-centric approach is that it makes sense to have special rules for communications infrastructure, because they are in a position to interfere with individuals' communications privacy at a different level than services that merely use the infrastructure. The main weakness of a service-centric approach is that such an approach can lead to, from a user's perspective, arbitrary differences between the privacy protections applicable to functionally similar services.

## D. A data-centric approach

39  A second approach to develop the scope of electronic communications privacy rules is data-centric. A data-centric approach protects privacy interests by setting rules for collecting and using types of personal data.[48] A data-centric approach to privacy regulation lies at the heart of at least a hundred data privacy laws around the world.[49] For instance, the general Data Protection Directive applies if "personal data" are processed.[50]

40  Another example of a data-centric approach to scoping rules is the stricter regime for "special categories" of personal data (also called sensitive data) in the Data Protection Directive. Special categories of data are defined as "personal data revealing racial or ethnic origin, political opinions, religious or philosophical beliefs, trade-union membership, and (...) data concerning health or sex life."[51] Processing such special categories of data is in principle prohibited, unless a legal exception applies such as medical necessity.[52] A member state can choose to allow data subjects to override this prohibition by giving their "explicit consent".[53]

41  At first glance, the e-Privacy Directive appears to follow a data-centric approach, regulating personal data, and providing a specific regime for location and traffic data. After all, article 3 states that the "Directive shall apply to the processing of personal data (...)." The provisions regarding traffic and location data particularise the general rules for personal data processing in the Data Protection Directive.[54]

42  However, a number of the e-Privacy Directive's provisions have a broader scope than setting rules for processing categories of personal data. For instance, article 1(1) clarifies that the directive gives "protection of the legitimate interests of subscribers who are legal persons," even though data related to legal persons generally do not qualify as personal data.[55] Similarly, article 5(3) of the e-Privacy Directive

---

---

applies, in short, to anyone that wishes to store or access information on a user's device, including if no personal data are involved. The provision applies to "information", and not to the narrower concept of personal data.[56]

**43** The e-Privacy Directive seems primarily concerned with protecting personal data in the electronic communications sector, but the relationship with the Data Protection Directive remains murky. As Rosier says about the e-Privacy Directive: "[i]t is (...) not always clear whether the exact scope of the terms in the provision should be determined only in the light of the definitions provided within the Directive or if it is also necessary to determine the scope of the terms in the Directive in the light of the provisions of the Data Protection Directive".[57]

**44** An advantage of a data-centric approach is that it can provide relative clarity. For example, general data protection law can be applied without engaging in open discussions about the scope or meaning of the right to privacy, a concept that is notoriously difficult to define. As De Hert & Gutwirth note: "[t]he strength of data protection (...) is not to be neglected. The complex question 'is this a privacy issue?' is replaced by a more neutral and objective question 'are personal data processed?'"[58]

**45** Even though a data-centric approach may offer relative clarity, the scope of the personal data definition still leads to debate, also in the context of electronic communications. For example, for behavioural targeting, companies often process individual but nameless profiles. Many behavioural targeting companies suggest that they only process "anonymous" data, and that, therefore, data protection law does not apply to them.[59]

**46** If personal data are within the scope of the special categories of data definition, the data controller must comply with stricter rules. For the data controller, this could be easier than assessing whether certain personal information is sensitive for a particular data subject in a particular context. At the same time, the question of whether certain data fall within the special categories of data definition can be difficult to answer. For instance, do location data revealing regular visits to specialised health clinics constitute medical data? Do images of people constitute special categories of data, because they can reveal race or ethnic origin?[60]

**47** As Ohm argues, an advantage of extra protection to certain sensitive data types is that the data types can provide a rule of thumb for a more nuanced approach that takes all relevant circumstances into account.[61] Simitis warns that a list of special data categories should be seen as "no more than a mere alarm device. It signals that the rules normally applicable to the processing of personal data may not secure adequate protection".[62]

**48** Considering what is at stake for users from a privacy perspective, it makes sense to single out traffic and location data for more strict regulation, as currently stipulated in the e-Privacy Directive. As the Advocate General of the European Court of Justice notes, traffic data are "in a sense more than personal."[63] Traffic data are "'special' personal data, the use of which may make it possible to create a both faithful and exhaustive map of a large portion of a person's conduct strictly forming part of his private life, or even a complete and accurate picture of his private identity."[64] Mobile devices basically function as location tracking devices, and communications metadata over longer periods can allow for a detailed mapping of an individual's social, professional, and private life, revealing many sensitive details.[65]

**49** Unfortunately, the current framework for traffic and location data has flaws. A key problem with the

---

existing rules for traffic and location data is that those rules only apply to electronic communications networks and services. Since many other parties, including mobile application providers also process such data, it seems questionable whether rules that only apply to electronic communications networks and services add value.[66]

**50** Furthermore, national data retention regimes break with the system of stricter rules for traffic and location data. The Data Retention Directive was declared invalid in a manner that leaves little room for Europe-wide blanket data retention.[67] Nonetheless, a number of member states have already adopted or proposed new data retention laws.[68] A review of the current e-Privacy Directive will have to address the question regarding which guarantees people should enjoy with respect to their electronic communications traffic and location data.

**51** The data-centric approach has weaknesses. By focusing solely on regulating personal data processing, the law may neglect the ultimate goal of protecting people and social welfare. As Bennet notes, the point at which certain information becomes "personal", information is increasingly difficult to determine. In addition, the rules for the fair processing of personal data can be insensitive to the means of extraction or capture of data. Additionally, even in the case that no personal data about a specific person are captured, there may still exist power imbalances that call for regulatory intervention.[69]

**52** For instance, occasionally people are shocked by the use of aggregated and anonymised data that escape data protection law.[70] To illustrate, the Dutch public reacted angrily when the police used aggregated information derived from data gathered by TomTom, a vendor of navigation and mapping products for cars. The police used the data to choose where to install speeding cameras.[71] The Dutch Data Protection Authority examined TomTom's practices, and from a data protection law perspective did not find significant issues.[72] The data obtained by the police were anonymised and aggregated, and thus outside the scope of data protection law.

**53** The TomTom case illustrates a broader problem of a data-centric approach in a world of "big data" analytics. Rules for processing personal data do not address the way in which processing other information, including aggregate statistical information based on personal data, can affect a person. Furthermore, anonymisation may take data outside the scope of data protection law, but does not guarantee that people are treated fairly.[73] Furthermore, as Gürses notes, anonymisation can even disempower the individual, when it is used to prevent people "from understanding, scrutinising, and questioning the ways in which these data sets are used to organise and affect their access to resources and connections to a networked world".[74]

**54** In addition, stricter rules for certain personal data types may not be nuanced enough. As Nissenbaum notes, sensitivity often depends on the context, rather than on the data type.[75] In 1976, Turn already argued: "[s]ensitivity is a highly subjective and context-dependent property of personal information – what one individual may consider very sensitive may be regarded with indifference by many others, and it is likely that there is a large range of sensitivity assessments for every information item." He adds: "[e]ven the same information item may be innocuous in one system of records, but very sensitive in another. For example, while a person's name is usually public information, it becomes sensitive when associated with a system of psychiatric treatment records".[76]

55  Such considerations lead a number of authors to criticise data protection law's stricter regime for special categories of data. McCullagh argues: "[t]he current approach of listing certain types of personal data as sensitive engages in an a priori classification exercise which is flawed. It is a fallacy. The privacy sensitivity of data cannot be pre-determined; rather it is influenced by contextual factors, and so, should be determined on a posterio basis".[77]

56  A final drawback of the existing rules for certain data categories is that the rules are seemingly based on the assumption that personal data will be generated. From a privacy perspective, it may make sense to consider the electronic communications architecture itself – and whether personal data need to be generated at all. Aiming to ensure that electronic communications networks and services are designed in a privacy-friendly manner could be more effective to protect privacy, than aiming to ensure that personal data are processed fairly after they have been generated.

57  While the data-centric approach has weaknesses, continuing the e-Privacy Directive's data centric approach has some merit. There is considerable experience with regulating personal data and with protecting privacy in electronic communications through rules for specific data types. Another argument in favour of a data-centric approach is that the e-Privacy Directive aims to complement and particularise the general data protection framework, which regulates the processing of personal data.

## E. A value-centric approach

58  A third approach to develop the scope of the electronic communications privacy framework is value-centric, and focuses on the fundamental societal values that need protection. These values include the right to private life and confidentiality of communications, as well as the fair processing of communication-related personal data, as protected through the Charter of Fundamental Rights of the European Union and the European Convention on Human Rights. As its preamble shows, the current e-Privacy Directive focuses on these fundamental values in the electronic communications context. The directive aims to protect the data protection and privacy rights from the Charter of Fundamental Rights of the European Union.[78]

59  In addition to protecting the confidentiality of private communications, the e-Privacy Directive provides for specific restrictions on the processing of communications related metadata, as discussed in the previous section. These specific protections should be seen in the light of the fundamental right to personal data protection, in the Charter of Fundamental Rights of the European Union.[79] Furthermore, electronic communications metadata fall within the scope of the right to private communications in the European Convention on Human Rights and the Charter of Fundamental Rights of the European Union.[80]

60  The way in which the protection of these values is operationalised in the e-Privacy Directive's provisions still resonates best with the use of electronic communications for telephone calls and other types of electronically mediated conversations between individuals, such as email. As mentioned in the introduction to this paper, we use electronic communications networks for many purposes, including shopping, distance-working, accessing news, and interacting with government. Therefore, the lawmaker should consider a wider range of privacy and communications related fundamental values at stake for individuals in contemporary electronic communications.[81]

61  Values that are currently underemphasised in the e-Privacy Directive are freedom of expression and the freedom to communicate more generally. These values are not specifically mentioned in the current e-Privacy Directive. However, the effective exercise of the right to freedom of expression is increasingly dependent on access to electronic communications networks, and on the conditions under which that access can take place, including the protection of privacy and personal data.[82]

62 Privacy and freedom of expression are closely related.[83] The early history of the right to confidentiality of communications illustrates the connection between that right and the right to freedom of expression. When the right to confidentiality of correspondence was developed in the late eighteenth century, it was seen as an auxiliary right to safeguard freedom of expression.[84] Nowadays the right to confidentiality of communications is primarily regarded as a privacy-related right, but the connection remains, as is illustrated in the Digital Rights Ireland judgment by the Court of Justice of the European Union, on the Data Retention Directive:

> It is not inconceivable that the retention of the data in question might have an effect on the use, by subscribers or registered users, of the means of communication covered by that directive and, consequently, on their exercise of the freedom of expression guaranteed by Article 11 of the Charter.[85]

63 A weakness of a value-centric approach to scoping electronic communications privacy rules is that the guidance from values could remain too vague. Most people would agree that we want to foster human dignity, freedom, equality and solidarity, as the EU Charter of Fundamentals Rights puts it.[86] Likewise, most would agree that living conditions should improve, and peace, liberty and democracy should be strengthened, as the preamble of the Data Protection Directive suggests.[87] However, operationalising such goals is difficult.

64 A distinction put forth by Bordewijk and Van Kaam of four communication models can help to operationalise a value-based approach.[88] Four types of communication can be distinguished: (i) the classic telecommunications model, including new forms of electronic correspondence; (ii) the consultation model, regarding access to information and electronically available resources; (iii) the registration model, regarding, for instance, tracking users for electronic marketing; (iv) the publishing model, regarding electronic publishing or broadcasting. Each communication model implicates different user interests and therefore calls for different types of privacy protection.

65 Regarding the classic telecommunications model, the rules for classic voice communication and data exchange of a similarly conversational nature in the e-Privacy Directive are the most advanced. The e-Privacy Directive focuses on protecting the privacy interests at stake in this model, including the confidentiality of communications and related traffic data. However, the e-Privacy Directive does not protect these interests for services that are functionally equivalent to telecommunication services. From a value-centric point of view, this situation is hard to defend. In addition, the underlying fundamental value of the freedom to communicate could be made more explicit in the e-Privacy Directive.

66 In the consultation model, people use electronic networks to access many informational resources, such as news, government information, medical records, educational offerings, and entertainment. This use of the network includes access to software that can be installed on the device and allows for new types of usage of the network. Here, the primary interest of the user is to access the network to enjoy these resources.

67 From a regulatory perspective, the question is under which conditions people should be able to access such resources. For instance, should it be possible to gain access to these resources without having to identify oneself or leaving an identifiable trace between different destinations? Additionally, considering the interests in societal inclusion and participation at stake with access to communications networks, should more specific provisions be adopted for the tracking and logging of network use? Currently, the e-Privacy Directive leaves access to information mostly unaddressed.

---

82   See Court of Justice of the EU 8 April 2014, C-293/12 (Digital Rights Ireland), par 28.

83   See Richards NM, 'Intellectual privacy' Texas Law Review 2008, Vol. 87, p.387; J.V.J. Van Hoboken, 'Search engine freedom: on the implications of the right to freedom of expression for the legal governance of search engines' (PhD thesis university of Amsterdam), Kluwer Law International 2012, p. 226.

84   See B.R. Ruiz, 'Privacy in telecommunications: a European and an American approach', Kluwer Law International 1997, p. 67; see also ECtHR 22 May 1990, Autronic AG v. Switzerland, par. 47.

85   Court of Justice of the EU 8 April 2014, C-293/12 (Digital Rights Ireland).

86   See the preamble of the Charter of Fundamentals Rights of the European Union.

87   Recital 1 of the Data Protection Directive.

88   See J.L. Bordewijk & B. van Kaam, 'Towards a new classification of tele-information services', InterMedia, 1986 Vol. 14, No. 1, pp. 16-21; J.C. Arnbak, J.J. Van Cuilenburg & E.J. Dommering, Verbinding en Ontvlechting in de Communicatie, een studie naar toekomstig overheidsbeleid voor de openbare elektronische informatievoorziening [Bundling and unbundling of communication, a study into future government policy regarding public electronic information provision], Cramwinckel 1990, pp. 7-9.

**68** The registration model concerns, for example, tracking users for electronic marketing. Electronic networks are often used to track, monitor, and reach users. Regarding such practices, the e-Privacy Directive offers some protection. First, current provisions regarding unsolicited communications aim to ensure that having an email address does not imply getting email from anyone that wants to reach a user.[89] Second, article 5(3) of the e-Privacy Directive applies to tracking users with cookies or similar technologies.

**69** The publishing model concerns electronic publishing or broadcasting. Electronic communication networks enable people to publish information and ideas, including information and ideas related to matters of public concern. There is a need to consider privacy guarantees that should apply to this kind of use of electronic communications. For instance, should the possibility to publish anonymously be protected?

**70** The current e-Privacy Directive does not contain rules that focus on the privacy protections connected to the use of the network for electronic publication and broadcasting purposes. Services outside the scope of the current electronic communications regulatory framework set the predominant conditions for publishing online. For instance, social media sites and other publication platforms can censor or remove individuals' contributions to online debate. Facebook sometimes censors users' posts.[90] But electronic communications networks and services could also interfere with freedom of expression. For example, access to specific publication platforms could be curtailed or compromised.[91] It may be necessary to adopt electronic communication privacy rules that protect users when they publish information for an online audience.

**71** In sum, a value-centric approach could help to highlight the freedom of communication as a core value in the electronic communications context. This approach can ensure a more systematic evaluation and robust protection of the privacy interests at stake in the communications sector. The four communication models discussed above can help to identify user interests that go beyond the privacy interests at stake in traditional telecommunications networks, but that do deserve to be protected in sector-specific electronic communications privacy laws.

# F. Conclusion

**72** In this paper, we introduce a distinction between three approaches to electronic communication privacy rules. The scope of the rules could be developed based on (i) a service-centric approach, (ii) a data-centric approach, and (iii) a value-centric approach. Each of the approaches has strengths and weaknesses. We provide the distinction between the three approaches as an analytical tool, to assist in discussions concerning the scope of electronic privacy rules. The recently announced review of the e-Privacy Directive provides an opportunity to improve the current scope.

**73** In a service-centric approach to develop the scope of electronic communications privacy rules, the scope of the rules is delineated on the basis of different services. The primary strength of the service-centric approach, which is currently dominant in the e-Privacy Directive, is clarity for market actors. The service-centric approach also conforms well to the infrastructural nature of electronic communications networks and services. The main weakness of a service-centric approach is that it can lead to different privacy rules for communication services that are functionally equivalent to users. The e-Privacy Directive should make clear that any party (rather than only "providers of publicly available electronic communications services" and "providers of public communications networks") must respect confidentiality of communications.

**74** A data-centric approach protects users' privacy interests through the proxy of regulating the processing of types of personal data. The data-centric approach is also present in the e-Privacy Directive, for instance in the rules for location and traffic data. We conclude that the protective rules for traffic and location data should be extended to information society services.

**75** Regulating the processing of certain data types can be a helpful proxy to protect user interests. Still, the rationale for having an electronic communications privacy framework should be protecting people – not data. A value-centric approach determines the scope of the rules based on the fundamental user interests at play. These interests go beyond the privacy interests at stake in traditional telecommunications networks. In particular, the lawmaker should more explicitly recognise freedom of expression and freedom of communication as fundamental values

---

89    Article 13 of the e-Privacy Directive.

90    See M. Heins, 'The Brave New World of Social Media Censorship', Harvard Law Review 2013,127 325.

91    See Citizen Lab and C. Anderson, 'The Need for Democratization of Digital Security Solutions to Ensure the Right to Freedom of Expression', Submission to the UN Special Rapporteur on the Promotion and Protection of the Right to Freedom of Opinion and Expression, 10 February, 2015, Appendix, p. 4, https://citizenlab.org/wp-content/uploads/2015/02/SR-FOE-submission.pdf (accessed 15 November 2015).

underlying the directive.

*   Dr. Joris van Hoboken is a Postdoctoral Research Fellow at New York University, School of Law, Information Law Institute (ILI). Dr. Frederik Zuiderveen Borgesius is a Researcher at the University of Amsterdam, Faculty of Law, Institute for Information law (IViR). The authors would like to thank the participants at the EuroCPR 2015 conference, and Achim Klabunde in particular, for their comments on an earlier draft of this paper. The authors also thank Nico van Eijk and the anonymous reviewer for their useful comments. Where this paper discusses article 5.1 and article 5.3 of the e-Privacy Directive, and the concept of personal data and special categories of personal data, the paper builds on, and includes sentences from, F.J. Zuiderveen Borgesius, Improving Privacy Protection in the area of Behavioural Targeting, Kluwer Law International 2015.